\documentclass{elsart}
 \usepackage{amsfonts}
\def \be {\begin{equation}}
\def \ee {\end{equation}}
\def \ba {\begin{eqnarray}}
\def \ea {\end{eqnarray}}
\newcommand {\dbar}{{d\kern-.22em\lower-.73ex\hbox{-}}}

\newcommand {\bk} {{\mathbf k}}

\newcommand {\bP} {{\mathbf P}}
\newcommand {\bp} {{\mathbf p}}

\newcommand {\bn} {{\mathbf n}}
\newcommand {\brr} {{\mathbf r}}

\newcommand {\bx} {{\mathbf x}}

\newcommand {\calC}{{\mathcal C}}

\newcommand {\calM}{{\mathcal M}}

\newcommand {\calO}{{\mathcal O}}
\newcommand {\calP}{{\mathcal P}}

 \newcommand {\calT} {{\mathcal T}}
 \newcommand {\calY} {{\mathcal Y}}
 \newcommand {\calZ} {{\mathcal Z}}

\begin{document}
\runauthor{PKU}
\begin{frontmatter}
\title{A general study on the volume dependence of spectral weights in lattice field theory}

 \author[PKU]{Zhi-Yuan Niu},
 \author[PKU]{Ming Gong},
 \author[PKUC]{Chuan Liu},
 \author[PKU]{Yan Shen}

 \address[PKU]{School of Physics, Peking University\\
               Beijing, 100871, P.~R.~China}
 \address[PKUC]{School of Physics and Center for High Energy Physics\\
               Peking University, Beijing, 100871, P.~R.~China}
                \thanks{Work supported in part by NSFC under grant No.10835002, No.10675005 and No.10721063.}

\begin{abstract}
 It has been suggested that the volume dependence of the spectral
 weight could be utilized to distinguish single and multi-particle states in
 Monte Carlo simulations. In a recent study using a solvable model,
 the Lee model, we found that this criteria is applicable
 only for stable particles and narrow resonances, not for the broad resonances.
 In this paper, the same question is addressed
 within the finite size formalism outlined by L\"uscher. Using a
 quantum mechanical scattering model, the conclusion that was found in previous Lee
 model study is recovered. Then, following similar arguments as in
 L\"uscher's, it is argued that the result is valid
 for a general massive quantum field theory under the same
 conditions as the L\"uscher's formulae. Using the spectral
 weight function, a possibility of extracting resonance parameters
 is also pointed out.
\end{abstract}


\end{frontmatter}

 \section{Introduction}
 \label{sec:intro}

 Low-energy hadron-hadron scattering plays an important role
 in the understanding of non-perturbative physics of strong
 interaction. Due to its genuine non-perturbative nature, such
 problems can only be studied from first principles using
 non-perturbative methods like lattice QCD.
 L\"uscher has outlined a finite-size formalism which enables
 us to calculate the elastic scattering phase shifts using lattice
 simulations~\cite{luscher86:finitea,luscher86:finiteb,luscher90:finite,luscher91:finitea,luscher91:finiteb}.
 Over the years, extensive numerical simulations have been carried out to the
 study on hadron-hadron scattering using L\"uscher's formalism, both
 within the quenched approximations and using gauge field
 configurations with dynamical quarks~\cite{gupta93:scat,fukugita95:scat,jlqcd99:scat,JLQCD02:pipi_length,chuan02:pipiI2,juge04:pipi_length,chuan04:pipi,ishizuka05:pipi_length,%
 CPPACS03:pipi_phase,CPPACS04:pipi_phase_unquench,savage06:pipi}.

 In lattice study on hadron spectroscopy and hadron-hadron scattering,
 the most important physical quantity is the energy of the system which is
 obtained via the measurements of various correlation functions.
 However, since a quantum field theory does not conserve
 particle numbers in general, the distinction between single- and multi-particle
 states becomes an important and delicate issue in lattice calculations.
 In the infinite volume, the difference is obvious since they have
 different kinematic behaviors: single-particle states
 have discrete energy eigenvalues when viewed in their rest frame
 while multi-particle states usually have
 continuous spectrum starting from the corresponding threshold. However,
 when performing a lattice simulation in a finite volume,
 all energy eigenvalues in the finite box become discrete.
 Therefore, other means have to to be applied in order to
 identify the particle nature of a corresponding state.

 In principle, differences between single- and
 multi-particle states still persist in a finite volume.
 For example, although both have discrete spectra, the level spacing between
 neighboring multi-particle scattering states becomes infinitesimally small
 while the level spacing between the neighboring single-particle states remains finite
 as the volume goes to infinity. However, it is difficult to utilize this difference
 as a practical criteria since this requires
 the computation of excited energy eigenvalues in Monte Carlo simulations
 which is usually quite challenging.
 Another method suggested by various authors is to
 use the so-called spectral weight as the identifier.
 This is the quantity which can be measured directly (and relatively easily) from
 Monte Carlo simulations, together with the corresponding energy eigenvalue.
 In a finite volume, the volume dependence of the
 spectral weight for a eigenstate is expected to show
 different behavior for single- and multi-particle states.
 For example, one expects the following empirical rule:
 the spectral weight to show little
 volume dependence for a single particle state (if properly normalized),
 while for a two-particle state, it is expected to show a $1/L^3$
 dependence where $L$ being the size of the cubic box.
 This expected difference in volume dependence can be measured
 in lattice simulations by performing the same calculation in two
 distinct volumes. As an example, this strategy has been used in Ref.~\cite{KFLiu04:pentaquark}
 to study the possible penta-quark state. Using this technique, the authors concluded
 that the expected penta-quark (single-particle) states measured
 in their lattice calculations are in fact kaon-nucleon two-particle scattering states.
 However, this conclusion is not so settled even in the first-principle lattice QCD
 calculations~\cite{Takahashi:2005uk,Csikor:2005xb,Ishii:2004qe}.
 Therefore, the volume dependence of the spectral weight indeed
 can provide us useful information about the particle nature of the
 corresponding state.

 In a previous model study, we have shown that the above mentioned
 empirical rule to distinguish single- and multi-particle states
 are in fact only valid for stable particles and narrow resonances.
 Using a solvable model, the Lee model, we showed that this rule
 is violated for broad resonances~\cite{chuan08:Lee_model}.
 A general formula for the spectral weight was obtained which
 can show either single- or two-particle volume behavior depending
 whether the width of the resonance is narrow or broad.

 In this paper, we attempt to
 generalize this conclusion that we obtained in the Lee model,
 to the case of general massive quantum field theory.
 For this purpose, the general L\"uscher's formalism is adopted.
 In previous studies, people have been focusing mainly on the
 energy eigenvalue (which directly enters the famous L\"uscher's formula)
 of the system within L\"uscher's formalism.
 However, since the spectral weight $W(E,L)$ of a given state
 is intimately related to the overlap of the exact energy eigenfunction
 with the free scattering states, we have to study the wavefunction
 of a energy eigenstate in a finite volume.
 In this paper, our study focuses on the wavefunction in  the $A^+_1$ sector and
 a formula for the spectral weight is thus obtained within the
 non-relativistic quantum mechanics model. By studying
 the volume dependence of the spectral weight in the large
 volume limit, we arrive at the same conclusion as we drew from
 the previous Lee model study. Then, following L\"uscher's arguments,
 this result is generalized to massive quantum field theory.
 Our results also show a possibility of extracting the resonance
 parameters from the spectral weight function on various volumes.

 This paper is organized as follows. In Sec.~\ref{sec:model}, we
 briefly review the quantum-mechanical model in the infinite volume.
 In Sec.~\ref{sec:model_torus}, the quantum-mechanical model is
 studied on a three-dimensional torus of size $L$. In this section,
 we derive the relevant formulae for the spectral weight function
 and study its volume dependence. It is found that similar
 conclusion is reached as in our previous study using the Lee model.
 We then argue that, under the same restrictions as in L\"uscher's
 formula, our results found in the quantum-mechanical model can
 be generalized to massive quantum field theory. The possibility
 of extracting resonance parameters from spectral weight is also
 discussed. In Sec.~\ref{sec:conclude}, we will conclude with some general
 remarks. Details on the evaluation of a function $F(k^2)$ are
 listed in the appendix.

 \section{The Model in the Infinite Volume}
 \label{sec:model}

 Consider a quantum mechanical model whose Hamiltonian
 is given by:
 \be
 \label{eq:hamiltonian}
 H= -{1\over 2m}\nabla^2+V(r)\;,
 \ee
 where the potential $V(r)$ is zero for $r>a$ with some $a>0$.
 We now discuss the energy eigenstates satisfying:
 $H\Psi(\brr)=E\Psi(\brr)$.
 One can expand the eigenfunction in terms spherical harmonics:
 \be
 \Psi(\brr)=\psi_{lm}(r)Y_{lm}(\bn)\;.
 \ee
 with: $\brr=r\bn$ and $\psi_{lm}(r)$ is
 the radial wave-function satisfying the radial Schr\"odinger
 equation:
 \be
 \label{eq:radial}
 \left({d^2\over dr^2}+{2\over r}{d\over dr}-{l(l+1)\over r^2}
 +k^2-2mV(r)\right)\psi_{lm}(r)=0\;.
 \ee
 where $E=k^2/(2m)$ being the energy eigenvalue of the state.
 It is well-known that, there exist only one solution to the radial
 Schr\"odinger equation that is bounded near the origin. This
 solution will be denoted as: $u_l(r;k)$. To fix the normalization,
 we impose the condition:
 \be
 \label{eq:normalize_u}
 \lim_{r\rightarrow 0} r^{-l}u_l(r;k)=1\;,
 \ee
 and the solution to the radial Schr\"odinger equation
 then has the form:
 \be
 \psi_{lm}(r)=b_{lm}u_l(r;k)\;,
 \ee
 with some constant $b_{lm}$ to be fixed by other conditions
 (normalization, boundary conditions, etc.).

 In the region $r>a$ where the interaction vanishes, the
 solution $u_l(r;k)$ are expanded in terms of spherical
 Bessel functions:
 \footnote{In this paper, we have adopted the same
 convention as in Ref.~\cite{luscher91:finitea} which  agrees with
 Messiah's book~\cite{messiah:book}.}
 \be
 \label{eq:ul_expand}
 u_l(r;k) = \alpha_l(k)j_l(kr) +\beta_l(k)n_l(kr)\;.
 \ee
 The coefficients $\alpha_l(k)$ and $\beta_l(k)$ have
 simple relation with the scattering phase shift:
 \be
 e^{2i\delta_l(k)}={\alpha_l(k)+i\beta_l(k)\over
 \alpha_l(k)-i\beta_l(k)}\;,
 \;\;
 \tan\delta_l(k)={\beta_l(k)\over\alpha_l(k)}\;.
 \ee
 In the low-energy limit: $k\rightarrow 0$, one normally defines:
 \be
 \alpha^0_l=\lim_{k\rightarrow 0} k^l\alpha_l(k)\;,
 \;\;
 \beta^0_l=\lim_{k\rightarrow 0} k^{-l-1}\beta_l(k)\;,
 \ee
 and the threshold parameters:
 \footnote{Assuming $\alpha^0_l\neq 0$ which is usually the case.}
 \be
 a_l\equiv {\beta^0_l\over \alpha^0_l}\;.
 \ee
 In particular, $a_0$ for $l=0$ is referred to as
 the $s$-wave scattering length. Other $a_l$'s for $l>0$
 are sometimes also called scattering lengths in
 the corresponding channel, although they do not
 have the dimension of a length
 \footnote{From normalization condition~(\ref{eq:normalize_u}), it is
 easy to verify that the spectral parameters $a_l$
 has the length dimension of $2l+1$.}.
 The threshold parameters $a_l$ are important because they
 characterize the behaviors in low-energy scattering processes.
 For example, we have:
 \be
 \delta_l(k) \simeq a_l k^{2l+1}+O(k^{2l+3})\;,
 \;\;
 (\mbox{\rm mod    } \pi)\;.
 \ee

 \section{The Model on a Torus}
 \label{sec:model_torus}

 We now enclose the system we discussed in the previous section
 in a large cubic box and impose the periodic boundary condition
 in all three spatial directions.
 The potential itself is also modified to $V_L(\brr)$ by periodically extending over the
 whole space: $V_L(\brr)=\sum_{\bn\in\mathbb{Z}^3}V(|\brr+\bn L|)$.
 For later convenience, we define the the so-called ``outer region" as:
 \be
 \Omega=\{\brr:\; |\brr+\bn L|>a,\;\; \mbox{for all } \bn\in\mathbb{Z}^3\}\;.
 \ee
 This is the region where the potential vanishes identically.
 We assume the size of the box is $L$ which is
 much larger than any of the physical scale in the system.
 In particular, we need to have $L \gg 2a$ so that the outer region
 admits free spherical wave solutions (asymptotic states).
 We now would like to study the change in the energy eigenvalues,
 the corresponding wave-functions and their possible connections with the scattering
 phase shifts in the infinite volume. Our discussion here will
 focus on the case of a cubic box whose relevant symmetry
 group being the cubic group $O(\mathbb{Z})$. Generalization to an arbitrary
 rectangular box can be performed easily by changing the symmetry
 group to the corresponding ones ($D_4$ or $D_2$, etc.).

 Since the boundary condition breaks rotational symmetry explicitly,
 we anticipate that energy eigenstates of the system will not have a definite
 angular momentum in general. To be specific, the original eigenstate in the
 $s$-wave will acquire mixtures from higher angular momentum
 modes (mainly $l=4$ for a cubic box).
 However, since the original radial wave-function $u_l(r;k)$ and the spherical
 harmonics forms a complete set in the functional space, we may
 still expand the true eigenfunction in the box in terms of them:
 \be
 \label{eq:eigenfunction_exact}
 \Psi(\brr;k)=\sum_{lm} b_{lm}u_{l}(r;k) Y_{lm}(\bn)\;.
 \ee
 where the coefficients are to be determined by boundary
 conditions and normalization.

 In the outer region $\Omega$, the solution are those singular, periodic
 solutions for the Helmholtz equation. Thus we may write:
 \be
 \Psi(\brr;k)|_{\brr\in\Omega}=\sum_{lm}v_{lm}G_{lm}(\brr;k^2)\;.
 \ee
 In the meantime, the outer solution can also be expanded
 in terms of spherical harmonics and the spherical Bessel
 functions $j_l(kr)$ and $n_l(kr)$:
 \be
 \label{eq:expand_Ylm}
 G_{lm}(\brr;k^2)={(-)^lk^{l+1}\over 4\pi}
 \left[Y_{lm}(\Omega_\brr)n_l(kr)
 +\sum_{l'm'}\calM_{lm;l'm'}Y_{l'm'}(\Omega_\brr)
 j_{l'}(kr)\right]\;,
 \ee
 The explicit expression for $\calM_{l'm';lm}(k^2_i)$ is given in
 Ref.~\cite{luscher91:finitea} which we quote here:
 \ba
 \label{eq:calM-zeta}
 \calM_{lm;js}(k^2) &=&
 \sum_{l'm'}{(-)^si^{j-l}\calZ_{l'm'}(1,q^2)\over
 \pi^{3/2}q^{l'+1}}
 \sqrt{(2l+1)(2l'+1)(2j+1)}
 \nonumber \\
 &\times &
 \left(\begin{array}{ccc}
 l & l' & j \\
 0 & 0  & 0 \end{array}
 \right)
 \left(\begin{array}{ccc}
 l & l' & j \\
 m & m' & -s \end{array}
 \right)\;.
 \ea
 Here we have used the Wigner's $3j$-symbols and
 $q=kL/(2\pi)$. The zeta function $\calZ_{lm}(s,q^2)$ is
 defined as:
 \be
 \label{eq:zeta_def}
 \calZ_{lm}(s,q^2)=
 \sum_{\bn} {\calY_{lm}(\bn) \over (\bn^2-q^2)^s}\;.
 \ee
 From the analytically continued formula,
 it is obvious from the symmetry of $O(\mathbb{Z})$
 that, for $l\leq 4$, the only non-vanishing zeta functions
 at $s=1$ are: $\calZ_{00}$, and $\calZ_{40}$. This is
 in accordance with the fact that $s$-wave and $g$-wave
 mixes with each other in a cubic box. In what follows, we will
 focus on the $s$-wave eigenfunction.

 \subsection{L\"uscher's formula in the $A^+_1$ sector revisited}

 In the remaining part of this paper, we will be only concerned with
 the energy eigen-functions in the $A^+_1$ sector, which is
 the analogue of $s$-wave in a cubic box.

 A good approximation for the $s$-wave dominated eigenfunction can
 be written as a superposition of $l=0$ and $l=4$ spherical
 harmonics with the $s$-wave component much larger than that
 of $g$-wave. To explicitly construct this type of wave-functions, we
 notice that the eigen-function in $A^+_1$ sector has to be
 invariant under cubic symmetries. It is easy to verify that,
 there are only two homogeneous harmonic polynomials which are invariant
 under cubic symmetry up to $l\le 4$.
 They can be conveniently expressed as:
 \be
 \calY_{00}={1\over\sqrt{4\pi}}\;,
 \;\;
 \calY_{40}+{\sqrt{70}\over 14}(\calY_{4,4}+\calY_{4,-4})
 ={15\over 4\sqrt{\pi}}\left(x^4+y^4+z^4-{3\over5} r^4\right)\;.
 \ee
 So, we may write the eigen-function in $A^+_1$ sector as:
 \be
 \label{eq:Psi_A1}
 \Psi^{(A^+_1)}(\brr;k) = b_{00}u_0(r;k)Y_{00}
 +b_{40}u_4(r;k)\left(Y_{40}
 +{\sqrt{70}\over 14}(Y_{4,4}+Y_{4,-4})\right)+\cdots\;,
 \ee
 with $|b_{40}|\ll b_{00}$ in the large volume limit.
 In other words, to ensure cubic symmetry, the general coefficients
 $b_{lm}$ at $l=4$ with different $m$ values must have definite ratios.
 In the outer region, using relation~(\ref{eq:ul_expand}), we have:
 \ba
 \label{eq:psi_in_out}
 \Psi^{(A^+_1)}(\brr;k)|_{\brr\in\Omega} &=&
  b_{00}[\alpha_0j_0(kr)+\beta_0n_0(kr)]Y_{00}(\Omega_\brr)
  \nonumber \\
 &+&b_{40}[\alpha_4j_4(kr)+\beta_4n_4(kr)]
 \left(Y_{40}+{\sqrt{70}\over 14}(Y_{4,4}+Y_{4,-4})\right)
 +\cdots\;.
 \ea

 On the other hand, we know that, in the outer region $\Omega$,
 the eigen-function can also be expanded into singular periodic solutions
 of Helmholtz equation.
 Since $G_{lm}\equiv \calY_{lm}(\nabla)G(\brr;k^2)$ with
 $G(\brr;k^2)$ being rotationally invariant, we see that
 in order to keep the eigen-function invariant under
 cubic symmetry, we must have the combination:
 $G_{40}+{\sqrt{70}/14}(G_{4,4}+G_{4,-4})$ in the expansion.
 Thus we may write:
 \footnote{For simplicity of the following equations,
 we have scaled out an overall factor $(4\pi/k)$ and
 an extra factor of $(1/k^4)$ for the coefficient
 of $G_{40}$.}
 \be
 \label{eq:psi_into_Glm}
 \Psi^{(A^+_1)}(\brr;k)|_{r\in\Omega}= \left({4\pi\over k}\right)
 v_{00}\left[G_{00}
 +{v_{40}\over k^4}\left(G_{40}
 +{\sqrt{70}\over 14}\left[G_{4,4}+G_{4,-4}\right]\right)
 +\cdots\right]\;.
 \ee
 The fact that such a combination respects cubic symmetry
 can also be checked explicitly.
 Using the expressions~(\ref{eq:expand_Ylm}) and
 (\ref{eq:calM-zeta}),
 we may write the expansion for $G_{00}$ as:
 \be
 \label{eq:G00}
 G_{00} = {k\over 4\pi}\left[
 (n_0+m_{00}j_0)Y_{00}+\sqrt{{7\over 12}}m_{04}j_4
 \left(Y_{40}+{\sqrt{70}\over 14}[Y_{44}+Y_{4,-4}]\right)
 \right]\;,
 \ee
 where we have introduced: $m_{00}=\calM_{00;00}$ and
 $m_{04}=2\sqrt{3/7}\calM_{40;00}$ for
 later convenience (see Ref.~\cite{luscher91:finitea} for the notation).
 Similarly, for the higher angular momentum functions, we have:
 \ba
 \label{eq:G_expand}
 G_{40} &=& {k^5\over 4\pi}\left[
 n_4Y_{40}+\calM_{40;00}j_0Y_{00}+\calM_{40;20}j_2Y_{20}\right.
 \nonumber \\
 &+&\left. \calM_{40;40}j_4Y_{40}+\calM_{40;44}j_4(Y_{44}+Y_{4,-4})
 \right]\;,\nonumber \\
 G_{4,4}+G_{4,-4} &=& {k^5\over 4\pi}\left[
 n_4(Y_{4,4}+Y_{4,-4})+2\calM_{44;00}j_0Y_{00}+2\calM_{44;20}j_2Y_{20}\right.
 \nonumber \\
 &+& 2\calM_{44;40}j_4Y_{40}+\left. (\calM_{44;4,-4}+\calM_{44;44})j_4(Y_{44}+Y_{4,-4})
 \right]\;,
 \ea
 In the above expansions, we have also utilized the
 following properties of the matrix elements
 $\calM_{lm;l'm'}$:
 \be
 \calM_{lm;l'm'}=\calM_{l'm';lm}=\calM_{l,-m;l',-m'}\;.
 \ee

 Note that in the expansion of $G_{40}$ and $G_{44}+G_{4,-4}$ in
 Eq.~(\ref{eq:G_expand}), there are terms with $l=2$, $m=0$ spherical harmonics.
 However, when we construct the combination
 $G_{40}+(\sqrt{70}/14)(G_{44}+G_{4,-4})$, the terms with $l=2$
 cancel out explicitly since: $\calM_{40;20}+(\sqrt{70}/7)\calM_{44;20}=0$
 which can be checked by looking into Table~E.1 in Ref.~\cite{luscher91:finitea}.
 Therefore we finally have:
 \ba
 \label{eq:G40}
 G_{40}+{\sqrt{70}\over 14}(G_{44}+G_{4,-4})&=&
 {k^5\over 4\pi}\left[\sqrt{{12\over 7}}m_{04}j_0Y_{00}\right.
 \nonumber \\
 &+&\left.
 (n_4+m_{44}j_4)\left(Y_{40}+{\sqrt{70}\over 14}[Y_{44}+Y_{4,-4}]\right)
 \right]
 \ea
 where $m_{44}=\calM_{40;40}+\cdots$. At this stage,
 it is worthwhile to point out that, $m_{00}$, $m_{04}$
 and $m_{44}$ that we introduced here are exactly those
 reduced matrix elements of $\calM$ in the $A^+_1$
 sector. Please refer to Ref.~\cite{luscher91:finitea} for
 further detailed explanations (especially Table E.1 and
 Table E.2 in the reference).

 Collecting relevant information from the expansions obtained thus far,
 i.e. Eq.~(\ref{eq:psi_into_Glm}), Eq.~(\ref{eq:G00})
 and Eq.~(\ref{eq:G40}), we have:
 \ba
 \label{eq:psi_out_out}
 &&\Psi^{(A^+_1)}(\brr;k)|_{r\in\Omega} =v_{00}
 \left[
  \left(n_0+m_{00}j_0+v_{40}\sqrt{{12\over 7}}m_{04}j_0\right)Y_{00}
  \right.\nonumber \\
 &+&\left.\left(\left[\sqrt{{7\over 12}}m_{04}+v_{40}m_{44}\right]j_4+v_{40}n_4\right)
 \left(Y_{40}+{\sqrt{70}\over 14}(Y_{4,4}+Y_{4,-4})\right) \right]+\cdots\;,
 \ea
 We should now match the two solutions given
 by Eq.~(\ref{eq:psi_in_out}) and Eq.~(\ref{eq:psi_out_out})
 in the outer region $\Omega$.
 This yields the following set of linear equations:
 \ba
 v_{00} &=& b_{00}\beta_0\;,\;\;
 v_{00}\left(m_{00}+\sqrt{{12\over 7}}v_{40}m_{04}\right)=b_{00}\alpha_0\;,
 \\
 v_{00}v_{40}&=&b_{40}\beta_4\;,\;\;
 v_{00}\left(\sqrt{{7\over 12}}m_{04}+v_{40}m_{44}\right)=b_{40}\alpha_4\;.
 \ea
 These four equations can be viewed as a set of homogeneous
 linear equations for the four coefficients: $v_{00}$,
 $b_{00}$, $v_{00}v_{40}$ and $b_{40}$. Demanding a
 non-trivial solution to exist requires the corresponding
 determinant of the $4\times 4$ matrix to vanish.
 Another simple way to proceed is to divide the second equation
 by the first and similarly divide the fourth one by the third.
 This will eliminate all coefficients except for $v_{40}$.
 We then arrive at:
 \be
 \cot\delta^{(0)}=m_{00}+\sqrt{{12\over 7}}v_{40}m_{04}\;,
 \;\;\;\;
 \cot\delta^{(4)}=m_{44}+\sqrt{{7\over 12}}m_{04}/v_{40}\;.
 \ee
 Eliminating $v_{40}$ from the above two equations then yields:
 \be
  \label{eq:match2}
 \left(\cot\delta^{(0)}-m_{00}\right)
 \left(\cot\delta^{(4)}-m_{44}\right)=
 m_{04}m_{04}\;.
 \ee
 This is exactly the equation obtained by general L\"uscher's
 method when we only consider the mixing between
 $l=0$ and $l=4$ waves~\cite{luscher91:finitea}.
 Therefore, using more explicit construction, not only
 have we recovered L\"uscher's formula, we also
 obtained an explicit approximate expression for the
 energy eigen-function in the $A^+_1$ channel which
 is given by Eq.~(\ref{eq:Psi_A1}) in general and
 given by Eq.~(\ref{eq:psi_into_Glm}) in the outer region.

 \subsection{The spectral weight function and its normalization}

 Now we would like to derive a formula for the spectral weight function which
 can be measured in a Monte Carlo simulation. Instead of working with
 general states, we will focus on the single- and two-particle states.
 These states naturally arise in the lattice study of hadron-hadron
 scattering and hadron spectrum.
 In such simulations, one constructs an operator (also known as the
 interpolating field operator), or operators if more than one
 is needed, within a specific symmetry sector of the theory.
 The correlation matrix among these operators are then computed
 by ensemble averaging over different gauge field configurations
 that is generated in a Monte Carlo simulation.

 For this purpose, we pass over to the second-quantized version of
 our quantum mechanical scattering model. In this model,
 two distinguishable particles scatter via a potential $V(r)$
 where $r$ being the distance between them. The center-of-mass coordinate
 of the two-particle system is separated
 out and the mass parameter $m$ in the Hamiltonian~(\ref{eq:hamiltonian})
 refers to the reduced mass of the two-particle system.
 For each type of particle, a local scalar field operator $\pi_i(\bx,t)$,
 with $i=1,2$ designating different types of particles,
 is introduced together with its momentum space counterpart:
 \footnote{For simplicity, we assume that the two particles
 are distinguishable.}
 \be
 \label{eq:fourier_def}
 \pi_i(\bx,t)={1\over \sqrt{L^3}}\sum_\bp \tilde{\pi}_i(\bp,t)e^{ i\bp\cdot\bx}
 \;,
 \tilde{\pi}_i(\bp,t)={1\over \sqrt{L^3}}\int d^3\bx \pi_i(\bx,t)e^{-i\bp\cdot\bx}
 \ee
 They satisfy the usual equal-time commutation relations:
 $[\pi_i(\bp,t),\pi^\dagger_j(\bk,t)]=\delta_{\bp\bk}\delta_{ij}$.
 Using free states made up of two particles, one from each type,
 one can form a state:
 \be
 |\Phi\rangle=\calO^\dagger(0)|0\rangle={1\over L^{3/2}}\sum_\bP \tilde{\Phi}(\bP)
 \tilde{\pi}^\dagger_1(\bP,0)\tilde{\pi}^\dagger_2(-\bP,0)
 |0\rangle\;,
 \ee
 with the interpolating operator $\calO(t)$ defined by:
 \be
 \calO(t)={1\over \sqrt{L^3}}\sum_\bP
 \tilde{\Phi}^*(\bP)\tilde{\pi}_1(\bP,t)\tilde{\pi}_2(-\bP,t)
 \;.
 \ee
 Requiring such a state to be normalized as:
 $\langle\Phi|\Phi\rangle=1$ yields the condition:
 \be
 {1\over L^3}\sum_\bP |\tilde{\Phi}(\bP)|^2=1\;.
 \ee
 If such a state were a bound state of two particles,
 $\tilde{\Phi}(\bP)$ would be the corresponding
 momentum-space wavefunction normalized according
 to the above equation.

 We can now define the corresponding correlation function:
 \be
 \calC(t)=\langle 0|\calO(t)\calO^\dagger(0)|0\rangle
 =\sum_E |\langle E|\calO^\dagger(0)|0\rangle|^2 e^{-Et}\;,
 \ee
 where $E$ and $|E\rangle$ represents the eigenvalue and
 eigenstate of the full Hamiltonian, respectively.
 By fitting the time-dependence of the correlation function obtained from
 Monte Carlo simulations, the exact eigenvalue $E$,
 and the corresponding spectral weight function $W(E)$,
 which is the coefficient in front of the exponential, is obtained.
 If we denote the overlap of two wavefunctions:
 \be
 \label{eq:overlap}
 O(E)=\langle E|\calO^\dagger(0)|0\rangle
 =\int d^3\brr_1d^3\brr_2\langle E|\brr_1,\brr_2\rangle\langle
 \brr_1,\brr_2|\calO^\dagger(0) |0\rangle
 \;,
 \ee
 the spectral weight function is simply given by:
 \be
 \label{eq:W_def}
 W(E)=|\langle E|\calO^\dagger(0)|0\rangle|^2=|O(E)|^2
 \;.
 \ee
 At this point, it is worthwhile to point out that the spectral
 weight function $W(E)$ defined  above depends explicitly on
 the normalization of $\calO$.

 Due to translational symmetry, the exact wave-function
 $\langle \brr_1,\brr_2|E \rangle$
 will only depend on the relative coordinate $\brr=\brr_2-\brr_1$.
 It is independent of the center-of-mass coordinate $\brr_c$.
 This means that, if the eigenstate $|E\rangle$ is normalized
 according to $\langle E|E\rangle =1$ as it should,
 the wave-function $\langle \brr_1,\brr_2|E \rangle\equiv\langle \brr |E\rangle$
 should be normalized according to:
 \be
 \label{eq:a1_normalize}
 \int_{\calT_3} d^3\brr |\langle \brr|E\rangle|^2=
 \int_{\calT_3} d^3\brr |\Psi^{(A^+_1)}(\brr;k)|^2={1\over L^3}\;.
 \ee
 Therefore, in order to compute the volume dependence of the spectral weight
 function, we first have to fix the normalization of $\Psi^{(A^+_1)}(\brr;k)$
 according to this convention.

 \subsection{Normalization of the energy eigenstates in $A^+_1$ sector}

 As discussed in the previous subsection,
 the wavefunction in the $A^+_1$ sector in Eq.~(\ref{eq:Psi_A1}) must be normalized
 properly on the torus $\calT_3$ according to
 Eq.~(\ref{eq:a1_normalize}). The integral of the eigen-function on the torus
 runs over two regions: the inner region where
 the explicit form of the wavefunction is not known
 and the outer region $\Omega$ where an approximate
 form of the function is given by Eq.~(\ref{eq:psi_into_Glm}).
 Although we do not know the exact form of the
 eigen-function in the inner region, we do know
 that the eigenfunction is bounded in this region.
 Since it is assumed that the interaction region
 is of size $a$ with $a\ll L$, therefore the
 integral in the normalization condition~(\ref{eq:a1_normalize})
 is dominated by the integral of the
 function in the outer region $\Omega$.
 Therefore, we may modify the normalization condition to:
 \be
  \int_{\Omega} d^3\brr |\Psi^{(A^+_1)}(\brr;k)|^2\simeq {1\over L^3}\;.
 \ee
 Since in the large volume limit, the eigen-function
 is dominated by the $s$-wave contribution, we may use
 the first term in Eq.~(\ref{eq:psi_into_Glm}) and write:
 \be
 \label{eq:norm_approx}
  \left({4\pi\over k}\right)^2|v_{00}|^2\left(\int_{\calT_3} d^3\brr |G_{00}(\brr;k)|^2
  -\int_B d^3\brr |G_{00}(\brr;k)|^2\right)\simeq {1\over L^3}\;,
 \ee
 where the second integral is over the interaction
 ball region: $B=\{\brr: r\le a,\; {\rm mod }\; L\}$.
 We now use the definition for $G_{00}$:
 \be
 \label{eq:G00_explicit}
 G_{00}(\brr;k)={1\over\sqrt{4\pi}L^3}\sum_\bp
 {e^{i\bp\cdot\brr}\over \bp^2-k^2}\;,
 \ee
 where the summation of $\bp=(2\pi/L)\bn$ is
 for all three-dimensional integers: $\bn\in \mathbb{Z}^3$.
 Substituting this expression into the first term and
 Eq.~(\ref{eq:psi_out_out}) into the second integral
 in Eq.~(\ref{eq:norm_approx}) we get:
 \be
  {k^2\over 16\pi^2|v_{00}|^2L^3}\simeq
 {1\over 4\pi L^3}\sum_{\bp}{1\over (\bp^2-k^2)^2}
 -
 {k^2\over 16\pi^2}
 \int^a_0 r^2dr \left(n_0(kr)+m_{00}j_0(kr)\right)^2\;.
 \ee
 The integral in the second term maybe evaluated directly within $r<a$.
 We thus obtain:
 \be
 \label{eq:v00_two_terms}
  {1\over |v_{00}|^2L^3}\simeq
 {4\pi\over k^2L^3}\sum_\bp{1\over (\bp^2-k^2)^2}
 -
 {a\over 2k^2\sin^2\Delta}\left[
 1-\left({\sin ka\over ka}\right)\cos(ka+2\Delta)
 \right]\;,
 \ee
 where we have utilized the definition:
 $m_{00}=\cot\Delta$.
 In the large volume limit, the first
 term on the r.h.s. of the above equation
 is much larger than the second (see appendix~\ref{app:F} for
 the explanation of this assertion).
 If we drop the second term, we then arrive at:
 \be
 \label{eq:v00_F}
 \left({4\pi\over k}\right)^2|v_{00}|^2L^3 \simeq
 4\pi \left({1\over L^3}\sum_\bp {1\over (\bp^2-k^2)^2}\right)^{-1}
 \equiv {4\pi\over F'(k^2)}\;,
 \ee
 where we have defined the function:
 \be
 F(k^2)={1\over L^3}\sum_\bp {f(\bp^2)\over \bp^2-k^2}\;,
 \ee
 where we have introduced a cutoff function $f(\bp^2)$ to
 regulate possible ultra-violet divergences. The property
 of this function in the large volume limit is addressed
 in appendix~\ref{app:F}. The relevant formula for us
 is given by Eq.~(\ref{eq:Fprime}).

 \subsection{Spectral weight in $A^+_1$ sector}

 We now evaluate the spectral weight using Eq.~(\ref{eq:W_def}) with
 the exact energy eigen-function given approximately by:
 $\Psi^{(A^+_1)}(\brr;k)\simeq (4\pi/k)v_{00}G_{00}(\brr;k)$.
 The overlap of the two wave-function is approximately given by:
 \be
 O=\left({4\pi\over k}\right)v^*_{00}{1\over\sqrt{4\pi L^3}}
 \sum_\bP {\tilde{\Phi}(\bP)\over \bP^2-k^2}\;.
 \ee
 Using the expression~(\ref{eq:v00_F}) and the expression
 in Eq.~(\ref{eq:Fprime}), we finally obtain
 $W(E)=|O|^2$ as:
 \be
 \label{eq:W_formula}
 W(E)={8\pi k |\varphi_L(k^2)|^2 \over \cot\delta_0(k)
 +{2\pi k^2\over \Delta \bp^2}\csc^2\delta_0(k)}
 ={8\pi k |\varphi_L(k^2)|^2 \over \cot\delta_0(k)
 +{2\pi E\over \Delta E}\csc^2\delta_0(k)}
 \;,
 \ee
 where the function $\varphi_L(k^2)$ is defined as:
 \be
 \varphi_L(k^2)={1\over L^3}\sum_\bP{\tilde{\Phi}(\bP)
 \over \bP^2-k^2}\;.
 \ee
 In the large volume limit, following similar derivation as
 in our discussion of function $F(k^2)$,
 this function goes over to:
 \be
 \varphi_\infty(k^2)=\calP\int{d^3\bp\over(2\pi)^3}
 {\tilde{\Phi}(\bp)\over \bp^2-k^2}+
 {k\tilde{\Phi}(k^2)\over 4\pi}\cot\delta_0(k)
 \;.
 \ee
 Thus the function $\varphi_L(k^2)$ has little volume dependence
 in the large volume limit.
 Therefore, the explicit volume dependence of the
 spectral weight function $W(E)$ comes mainly from the denominator
 in Eq.~(\ref{eq:W_formula}).
 Normally, if $\cot\delta_0(k)$ is not changing rapidly,
 the second term in the denominator of Eq.~(\ref{eq:W_formula}),
 which is proportional to $L^3$,
 dominates the result and one finds that the spectral weight
 is proportional to $1/L^3$. This is the typical two-particle
 spectral weight function. However, if there exists a rather
 narrow resonance at energy $E=E_\star$, then close
 to this resonance energy, one has approximately:
 \be
 \cot\delta(E)\simeq {E_\star-E\over \Gamma/2}\;,
 \ee
 where $\Gamma$ is the physical width of the resonance.
 In this case, we obtain:
 \be
 \label{eq:resonance}
 W(E)\simeq {4\pi k_\star\Gamma |\varphi(k^2_\star)|^2\over
 (E_\star-E)+\pi E_\star{\Gamma\over\Delta E}}
 \;.
 \ee
 If $\Gamma/\Delta E \ll 1$, then the quantity in the
 denominator is dominated by the first term and the
 spectral weight shows a typical single-particle behavior.
 This means that an extremely narrow resonance behaves like
 a stable particle. If on the other hand $\Gamma/\Delta E \gg 1$,
 which is always true for an extremely large volume (assuming
 the width of the resonance remains finite), the denominator is
 dominated by the second term and the spectral weight itself
 is roughly proportional to $1/L^3$ which is typical for
 a two-particle scattering state. We therefore arrive at the
 conclusion that the volume dependence of the spectral weight
 near a resonance is controlled by the ratio
 $(\Gamma/\Delta E)$.

 \subsection{Generalization to massive quantum field theory}

 Our results on the volume dependence of the spectral weight
 is obtained within a quantum mechanical model. In this subsection,
 we would like to generalize these results to massive quantum field
 theory, following the line of arguments in L\"uscher's
 formalism~\cite{luscher91:finiteb}.
 Using an effective Schr\"odinger equation (derived from the Bethe-Salpeper
 equation)~\cite{luscher86:finiteb}, L\"uscher has argued that,
 if the size of the box is large enough
 such that all quantum field theory effects are suppressed exponentially,
 the results obtained within the quantum-mechanical
 model can be carried over to the case of massive quantum field
 theory literally~\cite{luscher86:finiteb,luscher91:finiteb}.
 Here, we will assume that the same conditions
 are satisfied and thus our results obtained within the
 quantum-mechanical model are expected to be
 valid for massive quantum field theory.

 \subsection{Possibility of extracting the resonance parameters from the spectral weight}

 The relation established in Eq.~(\ref{eq:resonance}) opens up
 a possibility for extracting the width of a resonance
 if the spectral weight can be
 measured in the simulation. Assuming that there exists a single
 resonance in the energy region that we are interested in, and the
 contribution from this single resonance dominates the scattering,
 we simply rewrite Eq.~(\ref{eq:resonance}) as:
 \be
 {1\over W(E,L)}\simeq {1\over 4\pi k_\star |\varphi(k^2_\star)|^2}
 \left({E_\star-E \over \Gamma}
 +{\pi E_\star\over \Delta E}\right)\;.
 \ee
 Therefore, by fitting the function $1/W(E,L)$ for different
 $E$ and $L$ (hence different $\Delta E$ as well),
 it is possible to extract the width parameter $\Gamma$
 together with the resonance position $E_\star$ of the resonance.
 Note that in previous lattice calculations,
 focus has been mainly put on the energy levels, i.e. the values of $E$, only.
 No attention is paid to the associated spectral weight function $W(E,L)$
 which in fact can be obtained from the fitting procedure of
 the corresponding correlation functions {\em with almost no extra costs}.
 The study in this paper indicates that,
 the spectral weight function at various volumes also contains
 valuable information about the scattering and might also be
 utilized in some way. In fact, it can be used as an cross-check for the
 scattering phase obtained from the energy levels.
 Of course, this is only a possibility at this stage. The
 feasibility of this method has to be check in realistic simulations.

 \section{Conclusions}
 \label{sec:conclude}

 In this paper, we have studied the volume dependence of the
 spectral weight function which is accessible in Monte Carlo lattice
 simulations. Motivated by our previous study in the Lee model,
 it is expected that the spectral weight function shows little
 volume dependence for a stable or narrow resonance while for
 a broad resonance, it exhibits a typical $1/L^3$ dependence, the
 same as a two-particle scattering state.
 To verify this scenario, L\"uscher's formalism is adopted.
 It is first shown in a quantum mechanical model and then
 generalized to any massive quantum field theory, assuming that the polarization
 effects are exponentially suppressed following L\"uscher's arguments.
 In particular, we expect this scenario to be true also for
 QCD which governs the scattering of hadrons and therefore
 our result is relevant for lattice QCD simulations.
 Our final result for the spectral weight is summarized in
 Eq.~(\ref{eq:W_formula}) which exhibits either single-
 or two-particle volume dependence depending on the value
 of $\Gamma/\Delta E$ where $\Gamma$ is the physical width
 of the resonance and $\Delta E$ is the typical level spacing
 near the resonance in the finite volume. Possibilities of using
 the formula to extract the width of a resonance is also
 discussed.

 \section*{Acknowledgments}

 The author C.~Liu would like to thank Prof. K.F.~Liu from
 University of Kentucky, Dr. J.P.Ma from ITP, Academia Sinica,
 Dr. Y.~Chen from IHEP, Academia Sinica,
 Prof. Y.-B. Liu from Nankai University,
 Prof. J.-B. Zhang from Zhejiang University,
 Prof. H.~Q.~Zheng, Prof. S.~H.~Zhu and
 Prof. S.~L.~Zhu from Peking University for valuable discussions.
 \appendix

 \section{The function $F(k^2)$}
 \label{app:F}

 To study the normalization of the wavefunction
 $\Psi^{(A^+_1)}(\brr;k)$ in the large volume limit, we define the function:
 \be
 F(k^2)={1\over L^3}\sum_\bp {f(\bp^2)\over \bp^2-k^2}\;,
 \ee
 where we have introduced a cutoff function $f(\bp^2)$.
 The relevant function appearing in the normalization
 condition~(\ref{eq:v00_F}) is
 given by the derivative of $F(k^2)$ with respect to $k^2$:
 \be
 F'(k^2)={1\over L^3}\sum_\bp {f(\bp^2)\over (\bp^2-k^2)^2}\;.
 \ee
 We now follow the argument in Ref.~\cite{chuan08:Lee_model} to
 estimate the value of $F(k^2)$ for arbitrary value of $k^2$ in
 the large $L$ limit.
 We separate the summation into two parts with: $|\bp^2-k^2|\ge\epsilon$
 and $|\bp^2-k^2|<\epsilon$. The first part goes smoothly to the principle-valued
 integral $\phi(k^2)$ while the second summation may be written as:
 \ba
 {1\over L^3}\sum_{\bp,|\bp^2-k^2|<\epsilon}{1\over \bp^2-k^2}
 &=&{1\over L^3}\sum^{\infty}_{n=-\infty}{1\over
 \bp^2_\star+n\Delta\bp^2-k^2}\nonumber \\
 &=& -{\pi\over L^3\Delta\bp^2}
 \cot\left[\pi\left({k^2-\bp^2_\star\over\Delta\bp^2}\right)\right]
 \;,
 \ea
 where $\bp^2_\star$ is the value of $\bp^2$ that is closest to
 $k^2$; $\Delta\bp^2$ is the typical level spacing between
 neighboring $\bp^2$ values which can be estimated by:
 \be
 {L^3\over (2\pi)^3}2\pi\sqrt{\bp^2}\Delta\bp^2 =1
 \;\;\;
 \mapsto
 \;\;\;
 L^3\Delta\bp^2={(2\pi)^2\over \sqrt{\bp^2}}
 \ee
 Therefore we obtain:
 \be
 F(k^2)=\phi(k^2)-{k\over 4\pi}
 \cot\left[\pi\left({k^2-\bp^2_\star\over\Delta\bp^2}\right)\right]
 \;.
 \ee
 However, since it is easy to verify that:
 \be
 F(k^2)={\calZ_{00}(1;q^2)\over 2\pi^{3/2} L}
 \simeq {k\over 4\pi}\cot\delta_0(k)\;,
 \ee
 where we have utilized the approximate relation (L\"uscher's formula):
 \be
 \cot\delta_0(k)={\calZ_{00}(1;q^2)\over \pi^{3/2} q}\;.
 \ee
 We therefore seem to have: $\phi(k^2)=0$ in which case
 we recover the DeWitt's formula:
 \be
 \delta_0(k)=-\pi\left({k^2-\bp^2_\star\over\Delta\bp^2}\right)
 \;.
 \ee
 If one evaluate $\phi(k^2)$ explicitly, one gets:
 \be
 \phi(k^2)=\calP\int {d^3\bp\over (2\pi)^3}{1\over \bp^2-k^2}
 =4\pi\Lambda +2\pi k\ln\left|{\Lambda-k\over\Lambda+k}\right|\;,
 \ee
 with a sharp momentum cutoff $\Lambda$. This expression
 indeed goes to zero if we drop the constant term and
 taking $\Lambda\rightarrow\infty$.
 Consequently we have for the function $F'(k^2)$:
 \ba
 \label{eq:Fprime}
 F'(k^2) &=& -{1\over 8\pi k}\cot\left[\pi\left({k^2-\bp^2_\star\over\Delta\bp^2}\right)\right]+
 {k\over 4\Delta\bp^2}
 \csc^2\left[\pi\left({k^2-\bp^2_\star\over\Delta\bp^2}\right)\right]
 \nonumber \\
 &=& {1\over 8\pi k}\cot\delta_0(k)
 +{k\over 4\Delta\bp^2}\csc^2\delta_0(k)\;,
 \ea
 where in the second line we have used DeWitt's formula.
 Since $\Delta\bp^2\propto L^{-3}$, we find that
 $F'(k^2)\propto L^3$ in the large volume limit.
 This justifies the assertion made after
 Eq.~(\ref{eq:v00_two_terms}) in the main text.


 \end{document}